\documentclass[prl,twocolumn,a4paper,amsmath]{revtex4}
\usepackage[utf8]{inputenc}
\usepackage{graphicx}
\usepackage{amsmath}

\usepackage{color}
\begin{document}
\title{Averaging local structure to predict the dynamic propensity in supercooled liquids}

\author{Emanuele Boattini$^1$, Frank Smallenburg$^2$, Laura Filion$^1$}
\affiliation{$^1$Soft Condensed Matter, Debye Institute of Nanomaterials Science, Utrecht University, Utrecht, Netherlands \\
$^2$Universit\'e Paris-Saclay, CNRS, Laboratoire de Physique des Solides, 91405 Orsay, France}

\begin{abstract}
Predicting the local dynamics of supercooled liquids based purely on local structure is a key challenge in our quest for understanding glassy materials. Recent years have seen an explosion of methods for making such a prediction, often via the application of increasingly complex machine learning techniques. The best predictions so far have involved so-called Graph Neural Networks (GNN) whose accuracy comes at a cost of models that involve on the order of 10$^5$ fit parameters.  In this Letter, we propose that the key structural ingredient to the GNN method is its ability to consider not only the local structure around a central particle, but also averaged structural features centered around nearby particles. We demonstrate  that this insight can be exploited to design a significantly more efficient model that provides essentially the same predictive power at a fraction of the computational complexity (approximately $1000$ fit parameters), and demonstrate its success by fitting the dynamic propensity of Kob-Andersen and binary hard-sphere mixtures. We then use this to make predictions regarding the importance of radial and angular descriptors in the dynamics of both models.
\end{abstract}

\maketitle

Unraveling the interplay between structure and dynamics in glassy materials is a major challenge in condensed matter science. When a liquid is rapidly cooled down or compressed to the point where it almost turns into a glass, its dynamics slow down by many orders of magnitude, while its structure typically stays largely unchanged. The dynamics of such glassy fluids are heterogeneous, with some regions of particles rearranging much more rapidly than others \cite{ediger2000spatially, berthier2011dynamical}. 
Key to understanding this phenomenon is identifying structural characteristics that are associated with these heterogeneities \cite{royall2015role, tanaka2019revealing}.

Traditionally, correlating structure and dynamics relied on physical intuition: one can look for local structural motifs \cite{marin2020tetrahedrality,tong2018revealing, tong2019structural,malins2013identification, leocmach2012roles} -- such as icosahedra or tetrahedra  -- or for other (structure-dependent) local physical features of the system \cite{doliwa2003does, widmer2006predicting, widmer2008irreversible, richard2020predicting} -- such as local density, or potential energy   -- that are expected to play a key role in determining the dynamics. A novel approach was pioneered in 2015 by Cubuk {\it et al.} \cite{cubuk2015identifying}, who demonstrated that machine learning (ML) techniques could be trained to identify slow and fast regions in a glassy liquid. Since then, a number of works have demonstrated the power of both supervised and unsupervised machine learning for correlating structure and dynamics in a variety of glassy systems \cite{boattini2020autonomously,paret2020assessing,cubuk2015identifying,  schoenholz2016structural, schoenholz2017relationship, landes2020attractive}. 

In recent years, many studies attempting to unravel the link between structure and dynamics have tried to reveal structural quantities capable of predicting the dynamic propensity \cite{widmer2006predicting,hocky2014correlation, dunleavy2015mutual, pan2016correlation, tong2018revealing,  tanaka2019revealing, bapst2020unveiling, marin2020tetrahedrality, paret2020assessing, boattini2020autonomously, balbuena2021structural}. The dynamic propensity of a particle is the absolute \footnote{Note that some implementations of the dynamic propensity use the square distance, rather than the absolute one.} distance that the particle moves over a given time interval, averaged over many simulated trajectories starting from the same initial configuration. In essence, it provides a measure of the future mobility of a particle, as governed by the structure of its surroundings \cite{berthier2007structure}. To date, the most accurate predictions of the dynamic propensity in glassy systems were achieved by Bapst {\it et al.} \cite{bapst2020unveiling} using a highly advanced machine learning approach: Graph Neural Networks (GNN). While the GNN is quite accurate, its design philosophy draws on physical insights as little as possible, resulting in a high computational complexity, and making drawing physical insights from the fitted model more difficult.
Specifically, since the GNN designs its own structural descriptors, training it requires optimizing an enormous number of parameters ($\sim 70000$ in Ref. \cite{bapst2020unveiling}). As a result, the training requires both a large data set (to avoid overfitting) and significant computational effort. Nonetheless, the predictive power of GNNs clearly indicates that their model architecture is able to capture the essential physics required to predict dynamics from local structure. This raises two important questions that we address in this letter:  can we learn from  the  success  of  GNNs  what  structural  features  we need  to  consider  to  accurately  predict  local  dynamics, and can we exploit these observations to design a significantly more efficient model that performs equally well?

To address the first question, we need to consider the architecture of a GNN  and compare it to simpler machine learning approaches. In most ML approaches, such as the support vector machines (SVMs) used in Refs. \onlinecite{cubuk2015identifying, schoenholz2016structural}, the environment of each particle is captured via set of handcrafted structure functions centered around the particle under consideration, which provide information about e.g. the radial density profile and bond angle distribution. Subsequently, a model is fitted that relates these structural descriptors to a dynamical descriptor of each particle. In contrast, in the GNN used in Ref. \onlinecite{bapst2020unveiling}, the input is a graph, where each node represents a particle, and particles closer than a certain cutoff distance are connected by an edge which carries as information the vector connecting the two particles.  After an encoding step, this information is then passed through a number of recursive iterations. In each iteration, the graph is mapped to a new graph with the same topology, but with updated information on the nodes and edges, where the mappings are nonlinear functions described by neural networks. Finally, a ``decoder'' step is used to produce a prediction for the desired dynamical descriptor -- in the case of Ref. \onlinecite{bapst2020unveiling} the dynamic propensity. 

The key trick of the GNN lies in the fact that in each successive iteration, information from further away is incorporated into the information for a given node or edge -- in an average sense. Hence, in contrast to the more standard hand-crafted descriptors which are generally only centered on the particle under consideration, the GNN designs its own descriptors,  that can in principle take into account averaged structural features at significant distances away. As shown in the recent publication by Bapst {\it et al} \cite{bapst2020unveiling}, this strategy works very well at fitting the dynamics of a supercooled Kob-Andersen mixture, clearly outperforming all previous algorithms.

This raises the question:  would incorporating the shell-averaging concept of GNNs into handcrafted descriptors lead to similar predictive power? Here we design a set of descriptors that explicitly incorporate information from multiple neighbour shells and fit them to dynamical information using simple linear regression.

We begin with a set of descriptors that encode the structure around each particle $i$, denoted as the vector $\mathbf{X}_i^{(0)}$.  Note that many glassy systems consist of particles of two or more different species, and hence these descriptors take into account both the positions of the particles as well as their species.
For $\mathbf{X}_i^{(0)}$, we design a set of structural descriptors consisting of both radial and angular structure functions. For the radial descriptors, we consider the same type of functions that were used in Refs. \cite{schoenholz2016structural,bapst2020unveiling} in combination with SVMs. These functions essentially measure the density of particles at a distance $r$ from a reference particle $i$ in a shell of width $2\delta$, and are defined as follows:
\begin{equation}
    G_i^{(0)} (r,\delta,s) =  \sum_{j\neq i: s_j=s} e^{-\frac{(r_{ij}-r)^2}{2\delta^2}}
\end{equation}
where $i$ is the reference particle, $r_{ij}$ is the distance between particle $i$ and $j$, $s_j$ is the species of particle $j$, and $s$ is the species of particles whose density we wish to probe.
For the angular descriptors, inspired by standard bond-orientational-order parameters \cite{steinhardt1983bond}, we use an expansion of the local density in terms of spherical harmonics. First, for any given particle $i$, we define the complex quantities
\begin{equation}
    q_i^{(0)}(l,m,r,\delta) = \frac{1}{Z}\sum_{j\neq i} e^{-\frac{(r_{ij}-r)^2}{2\delta^2}}Y^m_l(\mathbf{r}_{ij}),
\end{equation}
where $Y^m_l(\mathbf{r}_{ij})$ are the spherical harmonics of order $l$, with $m$ an integer that runs from $m=-l$ to $m=+l$, and $Z = \sum_{j\neq i} e^{-\frac{(r_{ij}-r)^2}{2\delta^2}}$
is a normalization constant. We then construct a rotationally invariant local descriptor
\begin{equation}
q_i^{(0)}(l,r,\delta) = \sqrt{\frac{4\pi}{2l+1}\sum_{m=-l}^{l}| q_i^{(0)}(l,m,r,\delta)|^2}.
\end{equation}
The full vector $\mathbf{X}_i^{(0)}$ for a given particle $i$ then consists of the values of $G_i^{(0)} (r,\delta,s)$ and $q_i^{(0)}(l,r,\delta)$, evaluated for a fixed set of $r$, $\delta$, and $l$ as specified in the Supplemental Information (SI). 

In order to incorporate the shell-averaging concept from GNNs, we then introduce higher-order descriptors $\mathbf{X}_i^{(n)}$, where each consecutive $\mathbf{X}_i^{(n)}$ is defined as a local average of the previous order $\mathbf{X}_i^{(n-1)}$. Specifically:
\begin{equation}
\mathbf{X}_i^{(n)} = \frac{1}{C} \ \sum_{j:r_{ij}<r_c} e^{-r_{ij}/r_c}  \mathbf{X}_j^{(n-1)},
\end{equation}
where $r_c$ is a cutoff radius and $C = \sum_{j:r_{ij}<r_c}e^{-r_{ij}/r_c}$. Here, we choose $r_c$ to approximately correspond to the second minimum in the radial distribution function, and we have confirmed that the results are only weakly dependent on the exact value (see SI).

The total descriptors for particle $i$, which we denote $\mathbf{\mathcal{X}}^{(n_{\mathrm{max}})}$, is then the combination of descriptors (angular and radial) from each shell up to a maximum level of $n_{\mathrm{max}}$.

Now that we have introduced a new set of descriptors, we explore how well they can be used to fit the dynamics, and whether including multiple generations (i.e. larger $n_\mathrm{max}$) improves the fitting quality. For our fitting approach, we use linear regression including a regularization term, also known as Ridge regression. For this simple fitting algorithm, the number of fit parameters is simply equal to the number of descriptors (plus an offset), and the fit can be performed on a basic workstation in a matter of seconds.

As our dynamic quantity of interest, we use the dynamic propensity \cite{widmer2006predicting, berthier2007structure, tong2018revealing}, a standard method for quantifying the component of the dynamical heterogeneity that is encoded in the structure. For a given configuration, it is found by performing $M$  simulations of the same configuration, each initialized with a new set of velocities drawn from the Maxwell Boltzmann distribution. The dynamic propensity $d(t)$ of a chosen particle is then its average absolute displacement after time $t$. Note that in this Letter, all times are measured in units of the structural relaxation time $\tau_\alpha$.

\begin{figure}
\begin{center}
\includegraphics[width=0.8\linewidth]{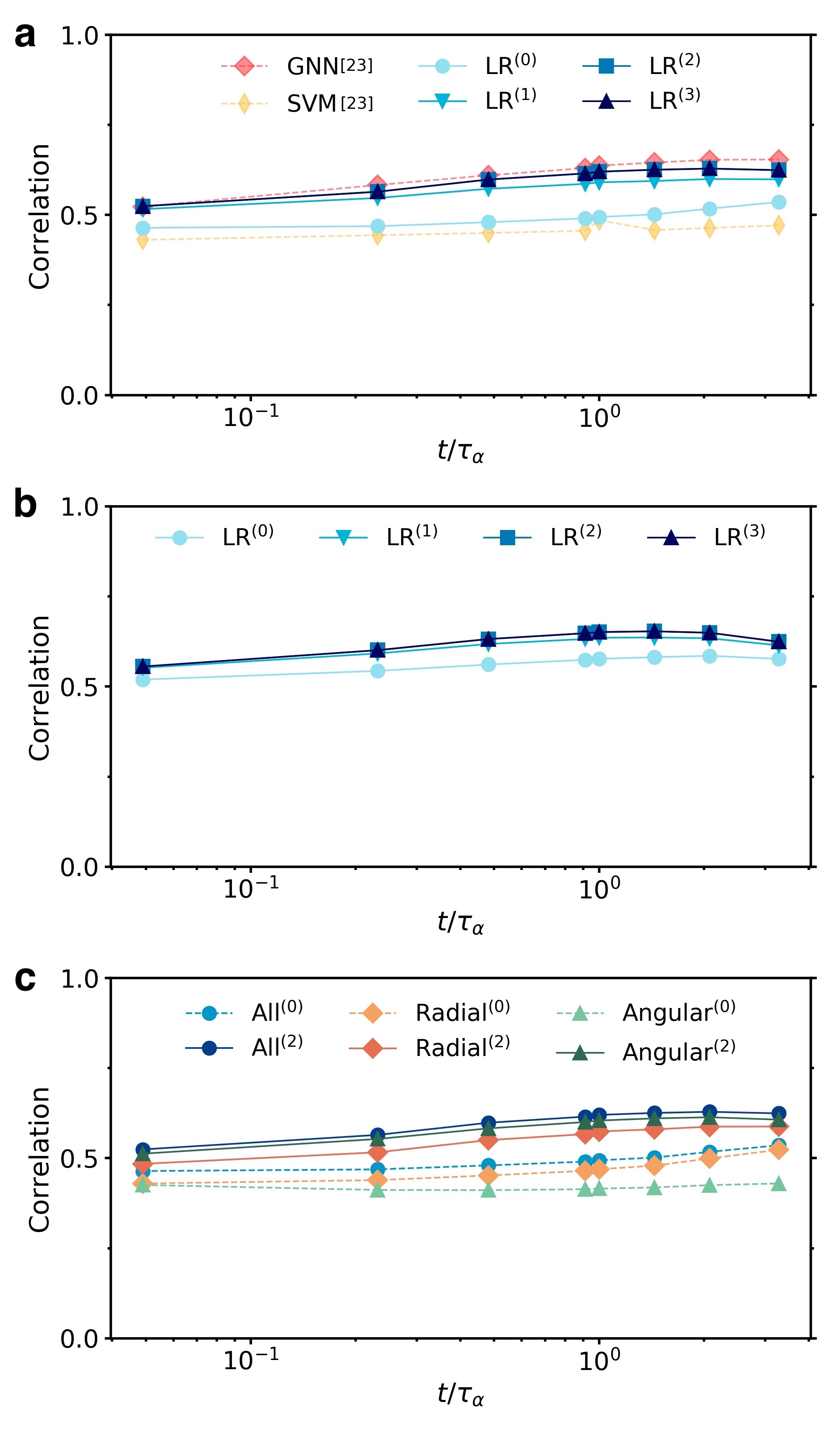}
\end{center}
\vspace{-0.5cm}
\caption{(a) Pearson correlation coefficient between predicted and actual propensities for the A particles of the KA system at temperature $k_B T / \epsilon_{AA} = 0.44$ and pressure $P \sigma^3_{AA}/\epsilon_{AA} = 2.93$. LR$^{(n_\mathrm{max})}$ refers to linear regression using structural descriptor up to order $n_\mathrm{max}$. For comparisons, the results obtained in Ref. \onlinecite{bapst2020unveiling} using support vector machines (SVMs) and graph neural networks (GNNs) are also shown. (b)Pearson correlation coefficient between predicted and actual propensities for the B particles of the same system. (c) Comparison of the Pearson correlation coefficient between predicted and actual propensities of A particles obtained using only radial descriptors, only angular descriptors, or both.}
\label{fig:KAall}
\vspace{-0.5cm}
\end{figure}

As our model system, we first focus on precisely the same system as Ref. \cite{bapst2020unveiling}, i.e. the Kob-Andersen (KA) mixture \cite{kob1995testing}, a well studied glass former that consists of an 80:20 mixture of non-additive Lennard-Jones particles (see SI for further details). We construct a set of descriptors consisting of 200 radial and 192 angular descriptors per generation. We begin by exploring how the number of averages (i.e. $n_\mathrm{max}$) influences our fit quality, as measured by the Pearson correlation coefficient  between the predicted and measured propensities (see SI). Clearly, as shown in Fig. \ref{fig:KAall}, for both species in the KA mixture, there is a significant improvement by including the first averaging  (LR$^{(1)}$), while adding a second averaging step (LR$^{(2)}$) only improves the correlations slightly. Adding a third averaging step does not lead to any further improvements.

The question is now: how does this prediction compare to other ML methods? As shown in Fig. \ref{fig:KAall}a,  the zero'th order descriptors lead to a prediction that is similar in quality to the SVM fit taken from Ref. \cite{bapst2020unveiling}. This is not surprising, as both the SVM method and our LR$^{(0)}$ prediction are linear models based on a similar set of descriptors. Much more interesting, however, is that our LR$^{(2)}$ results closely match the GNN results for all time scales.  This indicates that we have indeed incorporated in our averaged descriptors the same relevant structural features captured by the GNN.

One clear advantage of our approach compared to GNNs is the interpretability of the fitted model.  While GNNs approximate the propensity as a complex nonlinear function of the particles’ coordinates and species, our model is a linear combination of structural descriptors with a simple physical interpretation. Both the linearity of the model and the interpretability of the input descriptors can be exploited in order to unveil the structural features that are most relevant for predicting the dynamics. 
For example, we can ask the question: is radial (or density) information sufficient to accurately predict the dynamics, or do we also need angular information? This question is intriguing in the context of the Kob-Andersen system, as previous studies similar to our LR$^{(0)}$ model have identified radial information to be the most important \cite{paret2020assessing, schoenholz2016structural}. To answer this question, we separately fit the particles dynamic propensity using only the radial and angular descriptors, and show the results in Fig. \ref{fig:KAall}c.  When only non-averaged descriptors ($0$-th order) are included, we find the radial descriptors to be more informative on the dynamics (especially at long times), in agreement with  previous works \cite{paret2020assessing, schoenholz2016structural}. Interestingly, however, when averaged descriptors ($n_\mathrm{max}= 2$) are included, better predictions are obtained with the set of angular functions. One way to interpret this result is that when the region considered is small, the local density is the most important feature, whereas when considering larger length scales, the anisotropy of the structural environment becomes more relevant. This is consistent with the observation that the local environment of particles can show angular ordering over impressively long ranges in KA mixtures \cite{zhang2020revealing}.

\begin{figure}
\begin{center}
\includegraphics[width=0.8\linewidth]{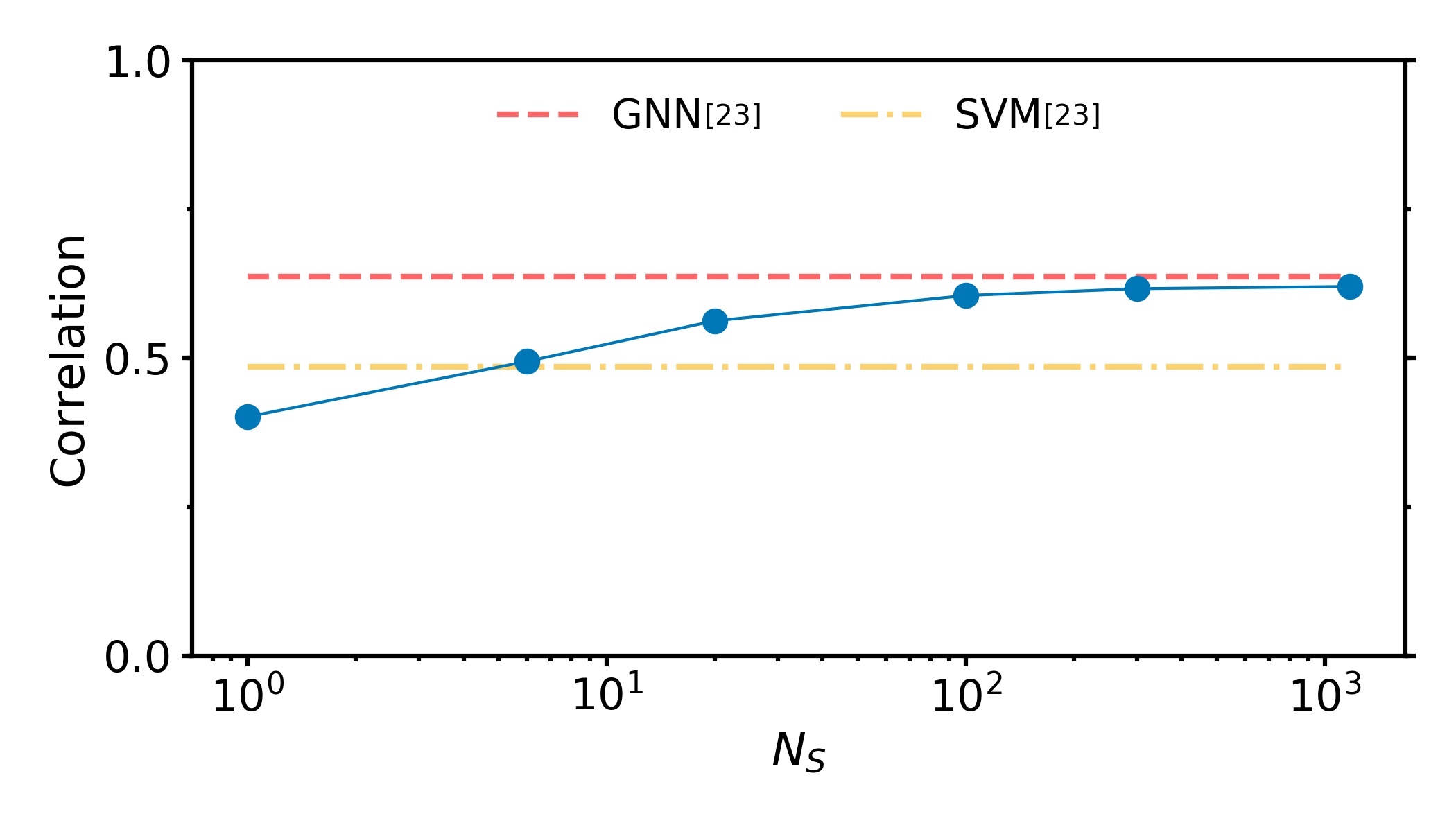}
\end{center}
\vspace{-0.5cm}
\caption{Pearson correlation coefficient between predicted and actual propensities of the A particles of the Kob-Andersen mixture (same state point as Fig. \ref{fig:KAall}) at $t=\tau_\alpha$ as a function of the number of selected descriptors. Lines represent the results obtained in Ref. \onlinecite{bapst2020unveiling} using SVMs and GNNs.} \label{fig:selection}
\vspace{-0.5cm}
\end{figure}

In our current approach we have included approximately $10^3$ descriptors. However, the linear nature of our model makes it easy to reduce the number significantly at a low cost in terms of accuracy -- a highly useful feature for future extensive explorations of the relationship between structure and dynamics of glassy fluids.  To this end, we employ the feature selection scheme introduced in Ref. \cite{boattini2020modeling} in the context of approximating many-body interactions. In this scheme, the most relevant descriptors are iteratively selected from a pool of candidates. At each step, the selected descriptor is the one that maximizes the linear correlation between the currently selected set and a target variable. The selection proceeds until the correlation stops increasing appreciably. Here, we consider the set of all descriptors up to order $n_\mathrm{max}=2$ as the pool of candidates, and use this scheme to select an optimal subset of $N_s$ descriptors for predicting the dynamic propensity at time $t$. In Fig. \ref{fig:selection}, we report the results of the predictions at $t=\tau_\alpha$ as a function of the number of selected descriptors. As shown in the figure, the best descriptor in the pool has a correlation of about 0.4 with the dynamic propensity. Moreover, using as few as $N_s=6$ descriptors, the results already exceed those obtained with SVMs in Ref. \cite{bapst2020unveiling} (that used 440 descriptors). After that, the results keep improving as $N_s$ increases, and do not change appreciably after $N_s\approx100$ descriptors have been selected. A table of the first 20 descriptors selected is given in the SI.

\begin{figure}
\begin{center}
\includegraphics[width=0.8\linewidth]{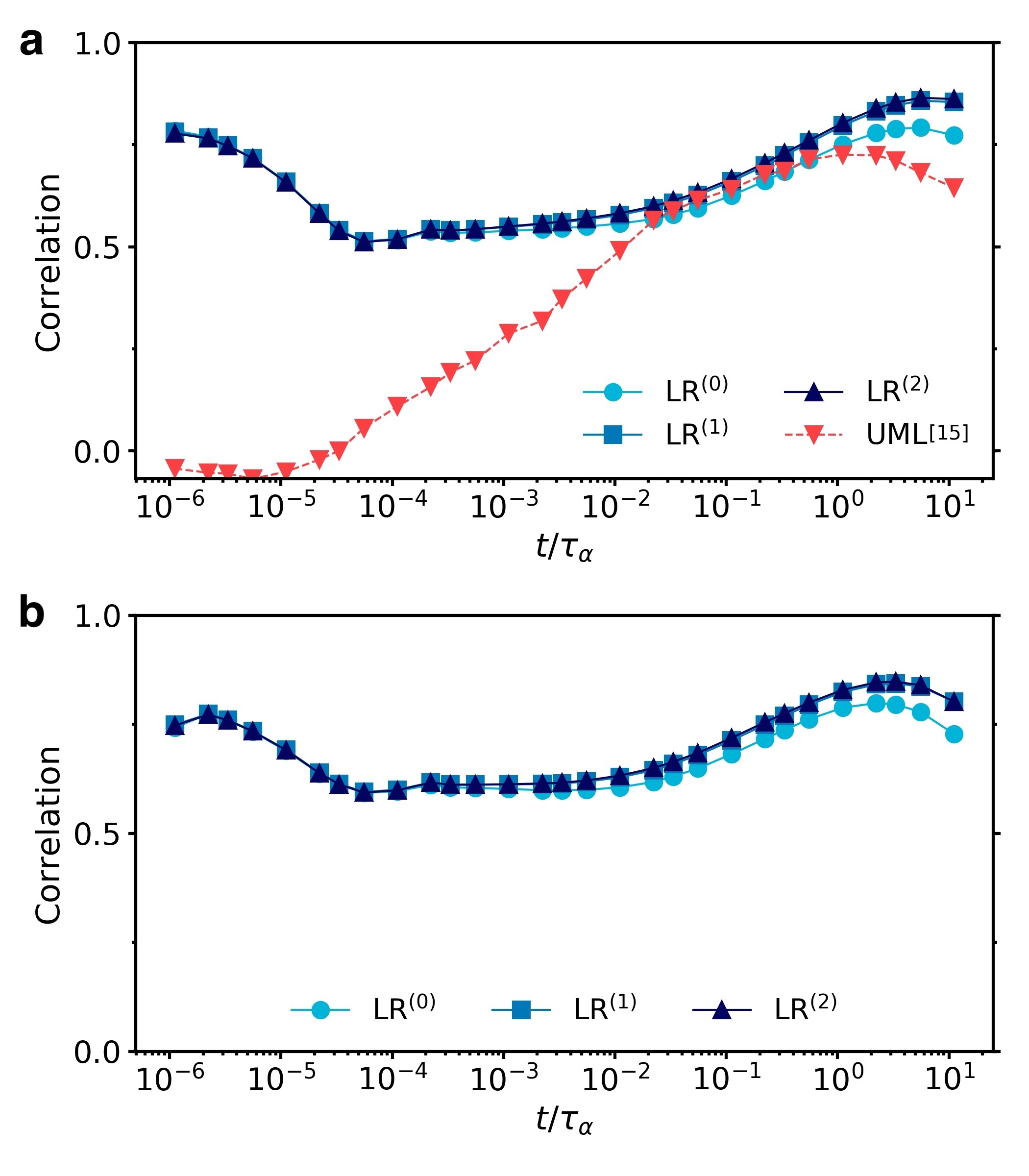}
\end{center}
\vspace{-0.5cm}
\caption{Pearson correlation coefficient between predicted and actual propensities for a hard-sphere mixture at size ratio $q = 0.85$, packing fraction $\eta = 0.58$ and composition $x_A = 0.3$. The results are shown for both (a) large A particles and (b) small B particles. For comparison, in (a) we also report the results of the UML approach from Ref. \onlinecite{boattini2020autonomously}. Note that these results were obtained on a different data set consisting of a single snapshot at the same state point considered here.} \label{fig:resultsHS}
\vspace{-0.5cm}
\end{figure}

The final question we would like to address is how robust this algorithm is -- e.g. how well does it perform on a different glass former? To this end, we consider a 30-70 binary mixture of large (A) and small (B) hard spheres with size ratio $\sigma_B/\sigma_A=0.85$ and packing fraction $\eta=0.58$. Previous works have shown that the local structure of this system correlates quite strongly with the dynamic propensity at times close to the relaxation time \cite{marin2020tetrahedrality,boattini2020autonomously}, with the strongest correlation reported associated with an unsupervised machine-learned (UML) order parameter based on averaged order parameters (somewhat comparable to  our LR$^{(1)}$ averaging). For hard spheres, we use the same set of descriptors as for KA, but omit radial descriptors taken at distances where no pairs of particles can exist (resulting in 172 radial descriptors). Using these descriptors with different choices of the maximum order $n_\mathrm{max}$, we fit the propensities at different times, and report the results obtained for both species of particles in Fig. \ref{fig:resultsHS}. Clearly, both the  LR$^{(1)}$ and LR$^{(2)}$ outperform the UML over all time frames examined, with the smallest difference appearing near $\tau_{\alpha}$. Interestingly, only approximatley 20 descriptors are necessary to reach approximately the optimum prediction accuracy (see SI, Fig. S3).  Similar to the KA system, the radial LR$^{(0)}$ descriptors outperform the angular descriptors at long time scales (see SI, Fig. S4). Overall, as seen in Fig. \ref{fig:resultsHS},  consistent with previous  observations \cite{boattini2020autonomously}, the predictions are much more accurate than those obtained for the KA system, likely due to the simpler dynamics of this system, which lacks both attractions and non-additivity. For the same reason, we also find that the inclusion of averaged descriptors has a weaker impact on the accuracy of the model, and the results essentially stop improving after including descriptors of order $n_\mathrm{max}=1$. 
An intriguing observation is the fact that the peak in correlation for LR$^{(1)}$ and LR$^{(2)}$ occurs at significantly longer time scales than the peak in correlation associated with the UML approach, and indeed at time scales several times longer than the structural relaxation time $\tau_\alpha$.

In conclusion, we have introduced a fast, easy to implement, linear-regression-based model for fitting dynamic propensities in glassy fluids from local structural descriptors.  Key to this model was the insight from GNNs that averaged structural features centered around nearby particles carry a significant amount of the necessary information required to predict the heterogeneous dynamics.  This observation enabled us to design a significantly more efficient model that provides essentially the same predictive power at a fraction of the computational complexity -- from the  $\sim 70000$ parameters of the GNN to approximately $1000$ parameters in the linear regression model at LR$^{(2)}$. Moreover, we show that by ranking the importance of the descriptors, we can further reduce the number of required descriptors by an order of magnitude. This result not only provides an efficient simple model for fitting the dynamic propensity of glassy fluids, but also suggests that similar local-average-based linear models should be considered in other situation where GNNs are applied to predict structural and dynamical properties of materials \cite{xie2018crystal,louis2020graph,schwarzer2019learning}.

Perhaps the most intriguing observation in this work is that the linear model presented here and the GNNs predict the dynamic propensity to essentially the same accuracy. Given that the dynamic propensity must be completely encoded in the structure by its definition, the new linear model opens the door to asking -- what structural information is missing to completely describe the dynamics? The linear model would appear to be missing information related to both 
anisotropic correlations within the averaged domains, as well as correlations between the averaged domains.  This observation should lay the foundation for further extensions and improvements in fitting the dynamics propensity in the future.

\section{Acknowledgements}
 We would like to thank Giuseppe Foffi for many useful discussions. L.F. and E.B. acknowledge funding from The Netherlands  Organisation  for  Scientific  Research  (NWO)  (Grant  No. 16DDS004), and L.F. acknowledges funding from NWO for a Vidi grant (Grant No. VI.VIDI.192.102).

\bibliography{ref}

\end{document}